\documentstyle[sprocl]{article}

\input{psfig}

\def\be{\begin{equation}}
\def\ee{\end{equation}}
\def\bea{\begin{eqnarray}}
\def\eea{\end{eqnarray}}

\begin{document}

\title{THE SUPERWORLD}

\author{NATHAN SEIBERG}

\address{School of Natural Sciences\\ Institute for Advanced Study\\ 
Princeton, NJ 08540, USA\\E-mail: seiberg@sns.ias.edu} 




\maketitle\abstracts{An overview of supersymmetry and its different
applications is presented.  We motivate supersymmetry in particle
physics.  We then explain how supersymmetry helps us analyze field
theories exactly, and what dynamical lessons these solutions teach us.
Finally, we describe how supersymmetry is used to derive exact results
in string theory.  These results have led to a revolution in our
understanding of the theory. }

\section{Introduction}

In this talk we will go on a guided tour through the superworld -- the
world of supersymmetric phenomena.  We will explain what supersymmetry
is and why many physicists expect to find it in the next generation of
experiments.  We will also show how powerful it is in leading to exact
results in field theory and in string theory and how these results
have revolutionized our understanding of string theory.

In section 2 we review the status of
the standard model of particle physics, its underlying principles and
its success.  We also review some of the flaws in the standard model,
which lead us to look for extensions of it.  In section 3 we introduce
supersymmetry as an extension of our ideas about the structure of
space and time.  In section 4 we explain why many physicist believe
that supersymmetry is likely to exist at low energies (around TeV) and
to be discovered soon.

In section 5 we turn to a different application of supersymmetry.  It
turns out that some aspects of supersymmetric field theories can be
analyzed exactly.  These are extremely complicated systems and the
fact that they can be analyzed exactly is by itself surprising.  More
importantly, the exact solutions exhibit interesting phenomena with
new lessons; among them is the crucial role played by
electric-magnetic duality in the dynamics.  The important applications
of supersymmetry to mathematics will not be reviewed here.

In the final section we turn to string theory and show how using the
magic of supersymmetry some nonperturbative information can be derived
in string theory.  This nonperturbative information has taught us many
new facts, completely changing our perspective on the theory.

It is logically possible that string theory does not describe Nature
and that supersymmetry will not be found in the TeV range.  Then, only
the applications in section 4 will survive.  It is also possible that
string theory describes Nature but supersymmetry is not present at low
energies.  Alternatively, it is also possible that string theory does
not describe Nature but supersymmetry is found soon.  My personal
prejudice is that we should get the whole ``package deal'' including
string theory and low energy supersymmetry.  Clearly, the fact that
supersymmetry naturally appears in one context makes it more likely
that it also appears in another.  It will be a shame if Nature does
not use a beautiful and powerful idea like supersymmetry.  However, as
physicists, we should never forget that only experiments are the final
judge about what constitutes a correct theory of Nature.

Many people have made crucial contributions to the developments of the
subject.  In order not to have a list of references longer than the
text, we omit all references.

\section{Review of the Standard Model}

The standard model of particle physics is based on the following
ingredients:

\begin{itemize}
\item
The theory respects special relativity.  In other words, space-time is
invariant under the 3+1 dimensional Poincar\'e symmetry.

\item
The theory is based on the principles of quantum mechanics.  It is
generally believed that the only quantum theory which respects special
relativity and is local (no action at a distance) and causal is local
quantum field theory.

\item
The theory has local gauge symmetry.  Unlike ordinary global
symmetries (like isospin) gauge symmetry allows arbitrary symmetry
transformation at different points in space-time.  Therefore the
symmetry group is really an infinite product of groups at different
space-time points.  Such a large symmetry group with arbitrary group
element at different space-time points is familiar from general
relativity and electrodynamics.  The specific gauge group of the
standard model is
$$SU(3) \times SU(2) \times U(1).$$ 
The gauge symmetry leads to gauge interactions which are mediated by
gauge particles.  For example, the electro-magnetic interactions are
mediated by photons.  Similarly, in the standard model we also have
gluons and $W$ and $Z$ gauge bosons.

\item
The matter particles in the standard model are in a representation of
$SU(3) \times SU(2) \times U(1)$.  These include the quarks, leptons
and Higgs boson.  Of these only the Higgs boson has not yet been
experimentally found ($\nu_\tau$ has been ``observed'' only
indirectly).

\item
The final ingredient of the standard model is its set of parameters.
These include the masses of the various particles, the fine structure
constant and a few others.  Of these only one parameter, the Higgs
mass, has not yet been measured.
\end{itemize}

The success of the standard model is unprecedented.  It is a fully
consistent theoretical theory.  Furthermore, there are many
experimental confirmations of the theory and there is no experiment
which is manifestly in contradiction with it.

Despite this spectacular success, it cannot be over stressed that this
is not ``The End of Science.''  In particular, all the ingredients of
the standard model are problematic:

\begin{itemize}

\item
We included special relativity but did not include general relativity
or gravity.  In Nature space-time is dynamical and can be curved.  In
the standard model, which does not include gravity, space-time is
static and provides a passive arena for the interactions.

\item
Trying to add gravity to the standard model and in particular to
combine general relativity with quantum mechanics leads to
contradictions.  Therefore, we must go beyond the framework of local
quantum field theory.

\item
Regarding the other ingredients of the standard model, we would like
to understand why the standard model is as it is.  Why is this the
gauge symmetry?  Why is this the particle spectrum?  Why are these the
values of the parameters?

\item
All the experimental tests of the standard model have been performed
at energies smaller than a few hundred GeVs.  Therefore, the standard
model should be viewed as an effective field theory valid up
to that energy scale.  At higher energies it can be extended to
another theory.  What is this theory?

\item
The last point allows for the possibility that there is no new physics
in the TeV range and new degrees of freedom show up only at much
higher energy, say $M_{\rm Planck} \sim 10^{19}_{\rm GeV}$, where
gravitational effects cannot be ignored and the theory must be
modified.  If this is the case, we face the hierarchy problem.  This
is essentially a problem of dimensional analysis.  Why is the
characteristic energy of the standard model, which is given by the
mass of the W boson $M_{\rm W} \sim 100 {\rm GeV}$ so much smaller
than the next scale, $M_{\rm Planck}$?  It should be stressed that in
quantum field theory this problem is not merely an aesthetic problem,
but it is also a serious technical problem.  Even if such a hierarchy
is present in some approximation, radiative corrections tend to
destroy it.  More explicitly, loop diagrams like the loop of
figure 1 restore dimensional analysis and move $M_{\rm W} \rightarrow
M_{\rm Planck}$.

\begin{figure}
\centerline{\psfig{figure=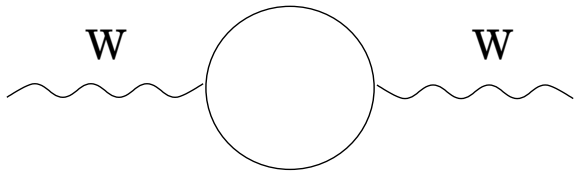,width=3in,angle=270}}
\caption[]{Loop diagram contributing to the mass of the W boson}
\label{fig1}
\end{figure}

\end{itemize}

\section{What is Supersymmetry?}

Supersymmetry is a new kind of symmetry relating bosons and
fermions.  According to supersymmetry every fermion is accompanied by
a bosonic superpartner.  For example, the quarks which are
fermions are accompanied by squarks which are bosons.  Similarly, the
gluons which are bosons are accompanied by gluinos which are
fermions. 

Another presentation of supersymmetry is based on the notion of {\it
superspace}.  We do not change the structure of space-time but we add
structure to it.  We start with four coordinates $X = t, x, y, z$ and
add four odd directions $\theta_{\alpha}$ $ (\alpha = 1, \cdots, 4)$.

These odd directions are fermions
$$\theta_{\alpha}\theta_{\beta}= -\theta_{\beta}\theta_{\alpha};$$
i.e.\ they are quantum dimensions and have no classical analog.
Therefore, it is difficult to visualize or to understand them
intuitively.  However, they can be treated formally.

The fact that the odd directions are anticommuting has important
consequences.  Consider a function of superspace
$$\Phi (X, \theta) = \phi (X) + \theta_{\alpha}
\psi_{\alpha}(X) + \cdots + \theta^{4}F(X).$$
Since $\theta_\alpha \theta_\alpha =0$, and since there are only four
different $\theta$s, the expansion in powers of
$\theta$ terminates at the fourth order.  Therefore, a function of
superspace includes only a finite number of functions of $X$.

Hence, we can replace any function of superspace $\Phi (X, \theta)$
with the component functions $\phi (X), \psi(X) \cdots$.  These
include bosons $\phi(X), \cdots$ and fermions $\psi(X), \cdots$.  This
facts relates this presentation of supersymmetry, which is
based on superspace, and the one at the beginning of this section,
which is based on pairing between bosons and fermions.

A supersymmetric theory looks like an ordinary theory with degrees of
freedom and interactions satisfying some symmetry requirements.  A
supersymmetric theory is a special case of more generic theory rather
than being a totally different kind of theory.

The fact that bosons and fermions come in pairs in supersymmetric
theories has important consequences.  In some loop diagrams, like in
figure 2, the bosons and the fermions cancel each other.

\begin{figure}
\centerline{\psfig{figure=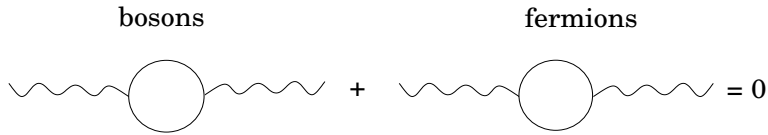,width=4in,angle=270}}
\caption[]{Boson-fermion cancellation in some loop diagrams}
\label{fig2}
\end{figure}

This boson-fermion cancellation is at the heart of most of the
applications of supersymmetry.  If supersymmetry is present in the TeV
range, this cancellation solves the gauge hierarchy problem.  Also,
this cancellation is one of the underlying reasons for being able to
analyze supersymmetric theories exactly.

\section{Supersymmetry in the TeV Range?}

There are several motivations for assuming that supersymmetry is
realized in the TeV range.  That means that the superpartners of all
the particles of the standard model have masses of the order of a few
TeV or less.

\begin{itemize}

\item
The main motivation is a solution of the hierarchy problem.  As we
mentioned in the previous section in supersymmetric theories some loop
diagrams vanish due to cancellations between bosons and fermions.  In
particular the loop diagram restoring dimensional analysis (figure 1)
is cancelled as in figure 2.  Therefore, in its simplest form
supersymmetry solves the technical aspects of the hierarchy problem.
More sophisticated ideas, known as dynamical supersymmetry breaking,
also solve the aesthetic problem.

\item
The second motivation for low energy supersymmetry comes from the
idea of gauge unification.  Recent experiments have yielded precise
determinations of the strength of the $SU(3)\times SU(2)\times U(1)$
gauge interactions -- the analog of the fine structure constant for
these interactions.  They are usually denoted by $\alpha_3$,
$\alpha_2$ and $\alpha_1$ for the three factors in $SU(3)\times
SU(2)\times U(1)$.  In quantum field theory these values depend on
the energy at which they are measured; i.e.\ these coupling constants
run.  The rates of change of these coupling constants depend on the
particle content of the theory.  Using the measured values of the
coupling constants and the particle content of the standard model, we
can extrapolate to higher energies and determine the coupling
constants there.  The result is that the three coupling constants do
not meet at the same point.  However, repeating this extrapolation
with the particles of the standard model and their superpartners the
three gauge coupling constants meet at a point (see figure 3).
How much weight should we assign to this result?  Two
lines must meet at a point.  Therefore, there are only two surprises
here.  First, the third line meets them at the same point -- there is
only one non-trivial number here.  Second, which is more qualitative,
the meeting point, the unification energy, is at a ``reasonable
value of the energy.''  My 
personal view is that this is far from a proof of low energy
supersymmetry but it is certainly encouraging circumstantial
evidence.

\begin{figure}
\centerline{\psfig{figure=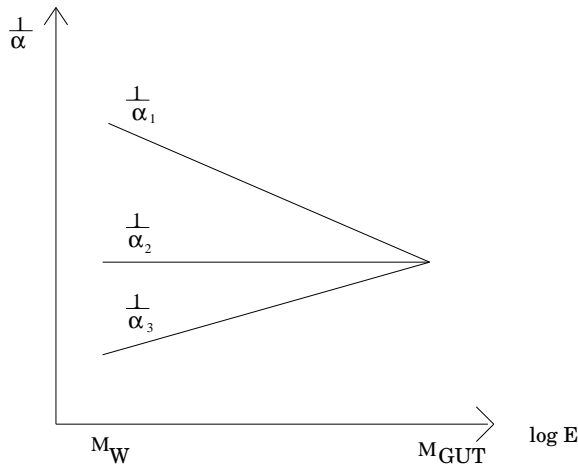,width=3in,angle=270}}
\caption[]{Coupling constant unification in supersymmetric theories}
\label{fig3}
\end{figure}

\item
The next generation of experiments at Fermilab and CERN will explore
the energy range where at least some of the superpartners are expected
to be found.  Therefore, in a few years we will know whether
supersymmetry exists at low energies.  If supersymmetry is discovered
in the TeV range, the parameters of the superpartners like their
masses and coupling constants will also be measured.  These numbers
will be extremely interesting as they will give us a window into the
physics of higher energies.

\item
Finally we should point out that some of these superpartners might
also be relevant for the dark matter of the Universe.

\end{itemize}

If supersymmetry is indeed discovered in the TeV range, this will
amount to the discovery of the new odd dimensions.  This will be a
major change in our view of space and time, comparable to and perhaps
bigger than the discovery of parity violation.  It should be stressed
that at the moment supersymmetry does not have a solid experimental
motivation.  If it is discovered, this will be one of the biggest
successes of theoretical physics -- predicting such a deep notion
without any experimental input!

\section{Exact Results in Quantum Field Theory}

Quantum field theory is notoriously complicated.  It is a non-linear
system of an infinite number of coupled degrees of freedom.
Therefore, (except in two dimensions) there are only a few exact
results in quantum field theory.  However, the special quantum field
theories which are supersymmetric can be analyzed exactly!

The main point is that these systems are very constrained.  The
dependence of some observables on the parameters of the problem is so
constrained that there is only one solution which satisfies all the
consistency conditions.  More technically, some observables vary
holomorphically (complex analytically) with the coupling constants
which are complex numbers.  Due to Cauchy's theorem, such analytic
functions are determined in terms of very little data: the
singularities and the asymptotic behavior.  Therefore, if
supersymmetry requires an observable to depend holomorphically on
the parameters and we know the singularities and the asymptotic
behavior, we can determine the exact answer.  The boson-fermion
cancellation, which we mentioned above in the context of the hierarchy
problem, can also be understood as a consequence of a constraint
following from holomorphy.

Another property of supersymmetric theories makes them tractable.
They have a family of inequivalent vacua.  To understand this fact
recall first the situation in a magnet.  It has a continuum of vacua,
labeled by the orientation of the spins.  These vacua are all
equivalent; i.e.\ the physical observables in one of these vacua are
exactly the same as in any other.  The reason is that these
vacua are related by a symmetry and the phenomenon of many vacua leads
to spontaneous symmetry breaking.

We now study a situation with inequivalent vacua.  Consider the case
of degrees of freedom with the potential $V(x,y)$ in figure 4.  The
vacua of the system correspond to the different points along the
valley of the potential, $y=0$ with arbitrary $x$.  However, as we
tried to make clear in figure 4,
these points are inequivalent -- there
is no symmetry which relates them.  More explicitly, the potential is
shallow around the origin but becomes steep for large $x$.  Such
``accidental degeneracy'' is usually lifted by quantum effects.  For
example, if the system corresponding to the potential in the figure
had no fermions, the zero point fluctuations around the different
vacua would have been different.  They would have led to a potential
along the valley pushing the minimum to the origin.

\begin{figure}
\centerline{\psfig{figure=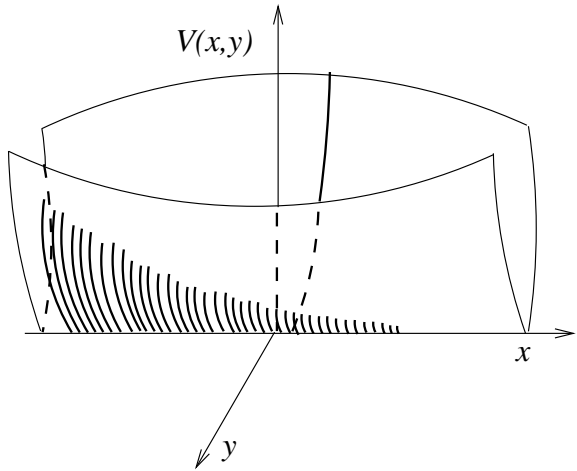,width=3in}}
\caption[]{Typical potential in supersymmetric theories exhibiting
``accidental vacuum degeneracy''}
\label{fig4}
\end{figure}

However, in a supersymmetric theory the zero point energy of the
fermions exactly cancels that of the bosons and the degeneracy is not
lifted.  The valleys persist in the full quantum theory.  Again, we
see the power of the boson-fermion cancellation.

We see that a supersymmetric system typically has a continuous family
of vacua.  It is referred to as {\it moduli space of vacua}, and the
modes of the system corresponding to motion along the valleys are
called moduli.

The analysis of the systems is usually simplified by the presence of
these manifolds of vacua.  Asymptotically, far along these flat
directions of the potential the analysis of the system is simple and
various approximate techniques are applicable.  Then by using the
asymptotic behavior along several such asymptotics as well as the
constraints from holomorphy the solution is unique.

This is a rather unusual situation in physics.  We perform approximate
calculations which are valid only in some regime and this gives us the
exact answer.  {\it This is a theorist's heaven -- exact results with
approximate methods!}

Once we know how to solve these system, we can analyze many examples.
The main lesson which was learned is the fundamental role played by
{\it electric-magnetic duality.}  It turns out to be the underlying
principle controlling the dynamics of these systems.

When faced with a complicated system with many coupled degrees of
freedom it is common in physics to look for weakly coupled variables
which capture most of the phenomena.  For example, in condensed matter
physics we formulate the problem at short distance in terms of
interacting electrons and nuclei.  The solution is the macroscopic
behavior of the matter and its possible phases.  It is described by
weakly coupled effective degrees of freedom.  Usually they are related
in a complicated, and in most cases unknown, way to the microscopic
variables.  Another example is hydrodynamics, where the microscopic
degrees of freedom are molecules and the long distance variables are
properties of a fluid which are described by partial differential
equations.

In one class of supersymmetric field theories the situation is similar
to that.  The long distance behavior is described by a set of weakly
coupled degrees of freedom.  As the characteristic length scale
becomes longer, the interactions between these effective degrees of
freedom become weaker, and the description in terms of them becomes
more accurate.  

In another class of examples there are no variables in terms of which
the long distance theory is simple.  The theory remains interacting
(it is in a non-trivial fixed point of the renormalization point).  In
these situations there are two (in some cases more than two)
descriptions of the physics leading to identical results for the
long distance interacting behavior.

In both classes of examples an explicit relation between the two sets
of variables is not known.  However, there are several reasons to
consider these pairs of descriptions as being electric-magnetic duals
of one another.  The original variables at short distance are referred
to as the electric degrees of freedom.  The other set of variables are
magnetic.

These two dual descriptions of the same theory give us a way to
address strong coupling problems.  When the electric variables are
strongly coupled, they fluctuate rapidly and their dynamics is
complicated.  However, the magnetic degrees of freedom are weakly
coupled.  They do not fluctuate rapidly and their dynamics is simple.
In the first class of examples the magnetic degrees of freedom are the
macroscopic ones.  They are massless bound states of the elementary
particles.  In the second class of examples there are two valid
descriptions of the long distance theory: the electric and the
magnetic ones.  As one of them becomes more strongly coupled, the
other becomes more weakly coupled.

Finally, using this electric-magnetic duality we can find a simple
description of complicated phenomena associated with the phase diagram
of the theories.  For example, as the electric degrees of freedom
become strongly coupled, they can lead to confinement.  In the
magnetic variables, this is simply the Higgs phenomenon
(superconductivity) which is easily understood in weak coupling.

We summarize the electric-magnetic relations in the following table:

\begin{table}[hb]
$$\vbox {\rm \halign {\strut#&\vrule#&\quad\hfil #\hfil\quad&\vrule#& 
\quad\hfil #\hfil\quad &\vrule#&\quad\hfil #\hfil\quad&\vrule#
\cr\noalign{\hrule} 
&&& &electric && magnetic & \cr\noalign{\hrule}
&&coupling & &strong & &weak & \cr\noalign{\hrule}
&&fluctuations && large && small & \cr\noalign{\hrule}
&&phase & &confinement && Higgs & \cr\noalign{\hrule}
}}$$
\end{table}

Apart from the ``practical'' application to solving quantum field
theories, the fact that a theory can be described either in terms of
electric or magnetic variables has deep consequences:

\begin{itemize}
\item
In theories of the first class of examples it is natural to describe
the magnetic degrees of freedom as composites of the elementary
electric ones.  The magnetic particles typically include massless
gauge particles reflecting a new magnetic gauge symmetry.  These
massless composite gauge particles are associated with a gauge
symmetry which is not present in the fundamental electric theory.
This is rather surprising because most people believed that such a
phenomenon cannot take place in four dimensions.  The lesson from
these examples is that {\it gauge invariance cannot be fundamental.}

\item
For theories of the second kind the notion of elementary particle
breaks down.  There is no invariant meaning to which degrees of
freedom are elementary and which are composite.  The magnetic degrees
of freedom are composites of electric ones and vice versa.  Again,
such behavior is very surprising in four dimensions.
\end{itemize}

\section{The String Revolution}

We do not know how to formulate string theory nor do we know its
underlying principles.  Surprisingly, this fact does not stop us from
making progress.  In particular, as in field theory, the magic of
supersymmetry allows us to obtain some exact results and to control
the theory in extreme situations.  These results have completely
changed our perspective on the theory.  In the remainder of this talk
we will briefly mention some of the main lessons:

\begin{itemize}
\item
Just as in supersymmetric field theories, string theory has many
inequivalent vacua -- a moduli space of vacua.  It turns out that the
supersymmetric compactifications of all five string theories are
connected.  A ``map'' of these vacua is given in figure
5.  At different boundaries of the map we find the five known string
theories as well as the mysterious eleven-dimensional theory whose low
energy limit is eleven-dimensional supergravity.  Without the magic of
supersymmetry only the vicinity of each boundary could be explored in
perturbation theory and there was no way to extrapolate from one
boundary to another.  Now, with these extrapolations, it is clear that
all the vacua are connected.  We conclude that instead of five string
theories there is only one theory with many solutions.  {\it The
theory is unique!}

\begin{figure}
\centerline{\psfig{figure=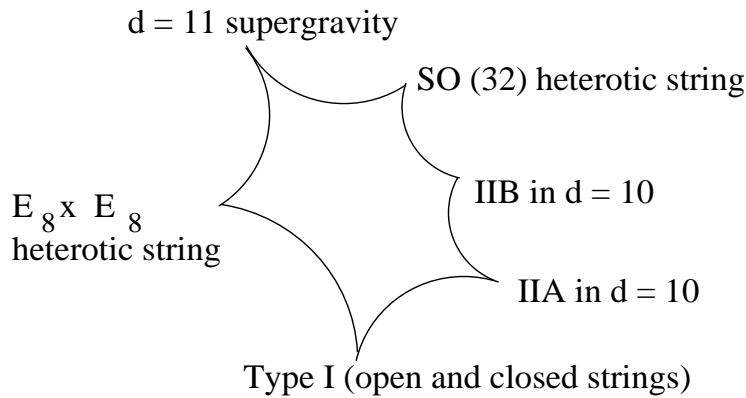,width=4in,angle=270}}
\caption[]{``Map'' of some of the vacua of string theory}
\label{fig5}
\end{figure}

\item
As we extrapolate from one boundary to another a phenomenon, which we
have already discussed in the previous section, takes place.  The
``elementary'' degrees of freedom at one boundary appear composite
elsewhere.  There is no universal object which appears elementary
everywhere and should therefore be viewed as preferred.  Furthermore,
in the boundary where the theory becomes eleven-dimensional there are
no strings at all.  We conclude that {\it the theory is not a theory
of strings.}  Therefore it is appropriate to change its name and it is
often being referred to as {\it M-theory} (M stands for magic,
mysterious, membrane, mother, ...).

\item
At various boundaries of the map in figure 5 there is a preferred
notion of space-time.  However, as we extrapolate from one boundary to
another, the underlying space-time becomes ambiguous.  The theory can
be described either as one kind of strings propagating on one
background or as another kind of strings propagating on another
background.  This ambiguity is known as {\it string duality}.  It has
led to a proposal to formulate the full theory in terms of the
dynamics of large matrices -- the coordinates of space-time are
non-commuting matrices in this approach.

\item
The map in figure 5 includes the value of a parameter which can
loosely be called $\hbar$.  As we approach various boundaries we seem
to take it to zero or infinity.  However, a more careful examination
of the theory shows that even as we set $\hbar \to 0$ the theory still
includes sectors, which remain quantum mechanical.  Furthermore, in
the eleven-dimensional vacuum there is no parameter like $\hbar$.  We
see that there is no classical theory whose quantization leads to
string theory.  Instead, {\it the theory is inherently quantum
mechanical!}

\item
Certain black-hole solutions of string theory where examined.  Using
the magic of supersymmetry an extrapolation from weak coupling to
strong coupling can be performed and one can exactly enumerate the
black-hole states.  It turns out to coincide with the number predicted
by the Bekenstein-Hawking entropy formula.  Therefore, the
black-hole entropy reflects the existence of many microscopic states.
This is a crucial step toward resolving the black-hole information
paradox.  It points in the direction that the full theory is unitary
and no information is being lost in Hawking radiation.
\end{itemize}

Unfortunately, these exciting developments have not yet led to direct
comparison with experiment.  The situations where exact answers are
possible are very idealized and have a lot of supersymmetry -- even
more than the amount of supersymmetry we expect to find in the TeV
range.  Even worse, before these developments one could have hoped that
the ten or eleven-dimensional vacua are somehow inconsistent.
Now, they appear perfectly consistent and are unified into a beautiful
picture.  Therefore, the question ``why don't we live in ten or eleven
dimensions?'' becomes sharper.

However, these developments are an enormous step toward uncovering
the underlying dynamical principles of string theory.

\section*{Acknowledgments}
This work was supported in part by DOE grants \#DE-FG02-90ER40542.
IASSNS-HEP-98/16.  Based on talks given at
various conferences and will appear in various proceedings.

\end{document}